\documentclass[aps,prb,preprint,showpacs,groupedaddress,endfloats]{revtex4}
\usepackage{graphicx}
\usepackage{setspace}
\usepackage{bm}

\newcommand{\be}{\begin{equation}}
\newcommand{\ee}{\end{equation}}
\newcommand{\bd}{\begin{displaymath}}
\newcommand{\ed}{\end{displaymath}}
\newcommand{\ba}{\begin{array}}
\newcommand{\ea}{\end{array}}
\newcommand{\bq}{\begin{eqnarray}}
\newcommand{\eq}{\end{eqnarray}}

\begin{document}
\begin{spacing}{2}
\title{Low energy electron scattering from DNA and RNA bases: shape resonances
and radiation damage}
\author{Stefano Tonzani}
\affiliation{JILA and Department of Chemistry,  University of Colorado, Boulder, Colorado 80309-0440 }
\author{Chris H. Greene}
\affiliation{Department of Physics and JILA, University of Colorado, Boulder, Colorado 80309-0440}

\date{\today}

\begin{abstract}
Calculations are carried out to determine elastic scattering cross sections and
resonance energies for
low energy electron impact on uracil and on each of the  DNA bases
(thymine, cytosine, adenine, guanine), for isolated
molecules in their equilibrium geometry. Our calculations are compared
with available theory and experiment. We also attempt to correlate this
information with
experimental dissociation patterns through an analysis of the
temporary anion structures that are  formed by electron capture in shape resonances. 
\end{abstract}
\pacs{}

\maketitle

\section{Introduction}
Reactions induced by electrons  drive nearly all the important chemical
processes in radiation chemistry, 
plasma etching in semiconductors, stability of waste repositories, and are also
fundamental in the dynamics of the atmosphere and interstellar clouds, with
processes such as dissociative recombination and electron attachment. 

In recent years, an
increasing importance has been recognized to these processes in biological
environments, especially in relation to radiation damage to nucleic acids (DNA
and RNA). These processes consist in the interaction of ionizing radiation
(like  $\alpha$, $\beta$ and $\gamma$-rays) with living tissue, generating
possibly mutagenic and carcinogenic byproducts, through a wide variety of
ionization, excitation and energy transfer processes, that can interest many
molecular species in the complex cell environment.

The important work of Sanche and coworkers
\cite{Sanche:DNA,Sanche:NIMP03,Sanche_review:MSR02} has shown that
damage to nucleic acids from ionizing radiation \cite{Mozejko_Sanche:REB03} (single and double strand
breaks in particular) can be generated through a mechanism involving low
energy electron attachment to the nucleic acid and subsequent bond breaking due to 
energy transfer to a vibrational mode of the 
temporary anion formed in the electron capture
step. These low-energy secondary electrons are generated by electron-impact ionization
caused by high energy electrons, originally produced directly by the ionizing radiation. In
the electron-impact ionization process, the scattered electron loses part of
its kinetic energy, while another electron is ejected, with energy much lower
than the first one.

In the past few years many studies have been devoted to understanding the
mechanism for the action of the low-energy electrons and their capability to
cause strand breaks.
\cite{Barrios:JPCB02,Schaefer:JACS05,Burrow:PRL04,Sanche_Burrow:PRL04,Burrow:JPCA98} A first general feature on
which there is wide agreement is that the electron capture is mainly due to the
DNA and RNA bases.
These molecules have extended aromatic systems, therefore there is a wide range
of low-lying unoccupied $\pi^*$ orbitals where an electron can be captured,
giving rise to a shape resonance, a temporary anion, in the range of energies
between 0 and 15 eV, where the experiments have found signatures of
electron-induced damage to nucleic acids.


The simplest of these bases are thymine, cytosine, uracil (pyrimidines,
monocyclic) and adenine and guanine (purines, bicyclic and generally larger than
pyrimidines). Their structures are shown in Fig. \ref{molec:structure}. 
In this paper we will present theoretical predictions of cross sections for elastic electron
scattering from these large molecules. Determination of the location, width,
and electronic structure of
resonances for a  single target molecule is an important step towards
understanding and possibly modeling the complex dynamics of 
DNA, which consists of multiple components. Specifically, besides the bases,
there are also the sugar
backbone, the phosphates, and also the solvation water,\cite{DNA:book} that
probably plays
a major role in stabilizing the temporary anions.
\cite{Sommerfeld:JPCA04,Gianturco:PRL04} No previous theoretical or experimental study of low-energy
electron scattering from all DNA bases is available for comparison (although a study at
intermediate energy has been carried out recently \cite{Mozejko_Sanche:REB03}), but our method has
proven its reliability in the study of small molecules \cite{Tonzani:JCP05} and
more extended systems like C$_{60}$, SF$_{6}$ and XeF$_6$.
\cite{Tonzani:unpublished}
 
Much experimental work also has been carried out on dissociative electron
attachment from DNA
bases,\cite{Sanche_Illenberger:JCP00,Illenberger_adenine:CPL03,Sanche_Burrow:PRL04} to understand what fragments are
generated.
We will discuss the possible connections with measured
dissociation branching ratios
that we can infer from the examination of the spatial shape and nodal surfaces
of the resonant wavefunctions.

\section{Theory}
\label{sec:Theory}
A detailed description of our method is available in Ref. 
\onlinecite{Tonzani:JCP05}. For this reason we will only review here the main points and
the changes we have implemented since that work, notably
a new polarization-correlation potential \cite{Lee_Yang_Parr:PRB88} which
permits enhanced predictive
capabilities.

The interaction between the $N$ electrons in a molecule and the scattered
electron is a many body problem that can be recast, \cite{Morr_Coll:PRA78} using the so called
static exchange approximation, into a one body problem for the continuum
electron with a nonlocal potential: \cite{Tonzani:JCP05}

\be
\label{1-body:eqn}
(-\frac{1}{2}\nabla^{2}+V_{s}-E)\phi_{0}(\vec{r})= \sum_{i,j}c_{i}c_{j}\sum_{k=1}^{N}
\phi_{ki}(\vec{r})\int{d\vec{r'}\frac{\phi^{*}_{kj}(\vec{r'})
\phi_{0}(\vec{r'})}{\mid\vec{r}
-\vec{r'}\mid}}
\ee
here $V_{s}$ is the electrostatic potential, $\phi_{0}$ is the scattered
electron orbital, while the other orbitals refer to target electrons, and the
$c_{i}$ are configuration interaction (CI) coefficients.
In this approximation only one state for the target is considered (the ground
state), whereby it is only suitable to describe electronically-elastic processes. The
nonlocal interaction consists of three main pieces: the direct electrostatic
contribution, the exchange term and the polarization-correlation potential. Of
these three, only the first is a local potential.

To describe electron scattering from a general, possibly very complicated
molecule, we use the R-matrix method, \cite{Greene:rev96} which partitions 
space into two regions: an internal region within which all the short-range physics is confined
and an outer region where only long-range interactions (like Coulomb or dipole
potentials) are important. Our calculation begins from a variational principle
\cite{Greene:rev96} for the logarithmic
derivative of the wave function at the boundary between the two regions

\be
\label{log:der}
{b\equiv - \frac{\partial{\log {(r\Psi)}}}{\partial r} = 2 \frac
{\int_V {\Psi^{*}(E-\hat{H} -\hat{L}) \Psi dV}}{\int_V{\Psi^{*}
\delta(r-r_0)\Psi dV}}}
\ee
where $\hat{L}$ is the Bloch operator, \cite{Tonzani:JCP05} and $r_{o}$ is the
boundary between the internal and external regions.
It is possible, after expanding the internal region wavefunction in a
suitable basis set, to recast the solution of Eq. \ref{log:der}
as an eigenvalue problem:

\be
\label{eigenvalue}
\underline{\Gamma}\vec{C}=({E-\underline{H}-\underline{L}}) \vec{C} =
\underline{\Lambda} \vec{C} b.
\ee

The external region is then treated  matching the solution of Eq.
\ref{eigenvalue} to the exact wavefunctions for the long-range tail of the
molecular potential. We show in Sec. \ref{sec:Dipole} how the contribution
from a long range dipole field can be included
in our method.
The basis set we use for the internal region of the $R$-matrix is a product of finite
element cubic polynomials in all three dimensions, using a grid in spherical
coordinates.

\subsection{Local Density Approximation (LDA)}
\label{sec:LDA}

To simplify further the description of our system we have to deal with the
nonlocality inherently present in the potential. To do this we use a local
density approximation for the exchange potential, which reduces it to a
functional only of the local density:
\be
\label{Pot:exch}
V_{ex}(\vec{r}) = -\frac{2}{\pi} k_{F} F(k_{F},E),
\ee
where $k_{F}$ is the local Fermi momentum:
\be
k_{F}(\vec{r})=(3 \pi ^{2} \rho(\vec{r}))^{1/3}
\ee
and $F$ is a functional of the energy and the local
density $\rho(\vec{r})$ (through the local Fermi momentum). The  functional
form we use for $F$ is called the Hara
exchange. \cite{Hara:69}  It has been extensively employed in continuum state
calculations, and it is energy-dependent.  The local exchange approximation, widely used also in
density functional calculations, \cite{Parr_Yang:book} has proven itself to give qualitatively correct
results, \cite{Tonzani:JCP05,Morr_Coll:PRA78} while being sufficiently simple
to implement computationally that it permits an exploration of complex
molecular species.

\subsection{DFT Polarization potential}
\label{sec:DFT}
We have recently added to our computer code the capability to use a parameter free
correlation-polarization potential,\cite{Gianturco:PRA93,Colle_Salvetti:JCP83} based on density
functional theory (DFT) ideas. As shown in Ref. \onlinecite{Lane:rev80} the
polarization-correlation contribution is physically related to the distortion-relaxation
effect on the molecule generated by the incoming electron. This is  extremely important for an accurate
description of the scattering process. The long range part of this potential
is
a simple multipole expansion, of which we retain only the induced dipole
polarization terms: 
\be
\label{polar:potential}
V_{pol} = -\frac{1}{2 r^{4}}(\alpha_{0}+\alpha_{2} P_{2}(\cos{\theta}))
\ee
where $\alpha_{0}$ and $\alpha_{2}$ are the totally symmetric and nontotally
symmetric components of the polarizability tensor, and are calculated $ab$ 
$initio$ using electronic structure codes.

In the volume where the electronic density of the target is not negligible,
this potential is nonlocal. The interaction can be approximated again as a
local potential, different forms of which have been suggested in the
literature.  The one
we use is based on DFT (specifically on the LYP potential of
Ref. \onlinecite{Lee_Yang_Parr:PRB88}) and it has yielded
reliable results in the work of Gianturco and coworkers. \cite{Gianturco:c60}
This form makes use of the electron density, its gradient and laplacian, which have
to be calculated for each target molecule. The short and long range potentials are
matched unambiguously at the innermost crossing point, whose radius is dependent on the
angles.

\subsection{Dipole physics}
\label{sec:Dipole}
Since the molecules we considered in this work have large electric dipole moments, there
is a need to consider the long range effect of the dipole field on the
scattered electron. Two possible options might be considered, either
extending the boundary of the R-matrix box far out to a region where the dipole
potential is very small, which would be extremely time-consuming for our
calculations, or matching to outer region functions adapted to the dipole interaction. We
choose this second route and, following the example of Clark,
\cite{Clark:PRA79} we define the matrix of the operator:

\be
(l^{2}-2 D \cos{\theta}) \Omega_{N}=N(N+1)\Omega_{N}
\label{Clark:operator}
\ee
where $l$ is the angular momentum operator, $\theta$ is the angle between the
incoming electron and the dipole direction, $D$ is the dipole moment, $N(N+1)$
and $\Omega_{N}$ are eigenvalues and eigenfunctions.  
We expand
$\Omega_{N}$ in a basis of spherical harmonics to diagonalize the
system in Eq. \ref{Clark:operator}. The
order of the spherical Bessel functions that are matched in the outer region
will be now $N$ (not an integer, in general) rather than the usual orbital
angular momentum quantum number $l$. Since the dipole moments
of the molecules in question are very large, 
 the dipole plus centrifugal potential becomes
attractive  for the first few channels which may thus have
a complex $N$. In such cases we define $N=-1/2+i\mu$ and the matching functions will become:\cite{Clark:PRA79}
\be
\bar{j}_{N}(k r)= \sqrt{\frac{\pi}{2r} } \frac{1}{\sinh{\frac{1}{2}\pi \mu}} Im(J_{i
\mu}(kr))
\label{Clark:mod_Bessel}
\ee
\be
\bar{n}_{N}(k r)= -\sqrt{\frac{\pi}{2r} } \frac{1}{\cosh{\frac{1}{2}\pi \mu}} Re(J_{i
\mu}(kr))
\label{Clark:mod_Bessel2}
\ee
where $J_{i\mu}$ is a cylindrical Bessel function. This allows
us to keep the functions in Eq. \ref{Clark:mod_Bessel} always real, and therefore have real
$K$-matrices. 

It should be mentioned, however, that at extremely low energies
these functions oscillate rapidly in energy as $\sin(\mu \log{kr})$, giving rise to $K$-matrices that
are not smooth functions of energy. Defining the base pair as in Ref. 
\onlinecite{Greene_Fano_Strinati:PRA79} solves the problem, but since we are not
interested in energies below about 0.5 eV in this study, the functions in Eqs.
\ref{Clark:mod_Bessel}-\ref{Clark:mod_Bessel2} will be sufficient.

The dipole plus centrifugal potential is attractive if the value of the dipole
moment is larger than a critical value ($D_{c}$=1.625 Debye for a nonrotating
dipole). In this case the dipole
interaction can bind the electron all by itself. In general, when rotation is included, the critical value of the dipole
moment to have a bound state is around\cite{Jordan:ANPC03} 2-2.5 D  and the number of
dipole-bound states is finite.
In the case of uracil, \cite{Ortiz:CPL99,Sommerfeld:JPCA04,Burrow:PRL04}
such a dipole-bound state exists at roughly 0.1 eV below the neutral ground
state energy, at the equilibrium geometry of the target molecule.
 
\subsection{Calculation details}
\label{sec:Details}
To calculate the target properties we have used the {\sc{GAUSSIAN 98}} program
suite, 
at the Hartree-Fock (HF) level of theory. We have noticed in the past
\cite{Tonzani:JCP05} that CI calculations
are in this case much more expensive and make comparatively little difference in the
final cross sections. All of the molecules we treat here are spin singlets in
their ground state.
For the scattering calculations we have used an IBM Power 4 supercomputer, each
calculation taking about 6 wall clock hours on 16 processors working in
parallel. The size of the matrices
generated is about 180000 by 180000 and the direct solution of the linear system
requires about 10 minutes per energy point when distributed over 16 processors.
The matrices are very sparse (fewer
than 0.5\% nonzero elements), and we use a direct sparse factorization method to
solve the linear system. The convergence of the calculation is such that
incrementing the number of sectors by 30\% lowers the energy of the resonances
by a further 0.1 eV in uracil; since it is already quite cumbersome to
carry out these calculations we deemed this level of convergence as sufficient for the
purposes of this study.
The geometry of the molecules is chosen to be the
equilibrium target geometry,  
optimized at the HF level with a 6-31G* basis set.

\section{Results}
\label{sec:Results}

To our knowledge there are no available experimental data or calculations of
low energy
electron scattering from the complete set of DNA bases. 
A study of electron attachment has been presented in Ref. \onlinecite{Burrow:JPCA98} and
the resonance positions are clearly marked. Compared to
these results, our calculations show resonances shifted typically by about 2 eV higher in
energy, but the energy spacing of the resonances is comparable to what is observed in
the experiment.Moreover 
the relative values of the widths of successive resonances resemble the
measured widths.
There is also a theoretical study
at intermediate energies, \cite{Mozejko_Sanche:REB03} and  calculations for
scattering from uracil; \cite{Gianturco:PRL04,Gianturco:uracil_JCP04} in
the following  
we compare these results to ours.

We have already mentioned that the heterocyclic DNA bases have many low-lying
unoccupied orbitals, so it is not surprising that their elastic cross sections
for electron scattering exhibit  many shape resonances. These may be viewed as a
capture of the scattered electron into one of these antibonding orbitals to form
a short lived negative ion state.\cite{Dehmer:res_JCP81,Dehmer:res_ACS84}

Since all these molecules have, in their equilibrium configuration,  
only one symmetry element - reflection through the molecular plane - we will characterize the resonances
as being of $\sigma$ type (no node in the plane) or $\pi$ type  (when they have instead a
node in the plane) rather than using the $A'$ and $A''$ labels as is customary for the $C_{s}$ group.

\subsection{Positions and widths of resonances}
\label{sec:Cross_sections}
A general comparison of partial  elastic cross sections for all five of these molecules 
is shown in Fig. \ref{fig:total_cross_section}, while in the following we
give a more detailed description and compare with information available in the
literature. 
Also, a plot of total time-delays (see also Sec. \ref{sec:Resonances} for
details)  is provided in Fig.
\ref{fig:deriv_eigenphase_sum} to show the resonances is more detail.

Since we are dealing with polar molecules, applying the fixed-nuclei
approximation as it stands makes the partial wave expansion of the forward scattering
amplitude divergent. Due to the long-range nature of the dipole
interaction, in fact, all partial waves would contribute to the scattering
process, causing an infinite scattering in the forward direction and therefore
infinite integral cross sections. There is a method, extensively discussed in the
literature, \cite{Gianturco:water_JCP98,Clark:PRA77} to deal with this problem by
means of a Born closure formula, which yields a finite integral cross section
once 
molecular rotations are included. We will not
pursue this further, since existing
experiments are not likely to deal with such detailed rotational structures.
Therefore our
cross sections and time-delays include only up to $l_{max}=10$ and omit all higher
partial waves.  The correction would be proportional  to the
dipole moment and inversely proportional to the smallest rotational spacing. For the
DNA bases the dipole
moment is large, while the rotational spacing is
small. 
Therefore the correction can be quite large
especially at very low energy. The correction would thus tend to mask the resonant
structures, which are the most interesting observables and which have been measured in
experiments.
All of the calculated cross sections grow rapidly when the incident electron
energy decreases below 1 eV, which is a
signature of the role played by the dipole field in pulling in the electron and
which is very common in electron scattering from polar
molecules. \cite{Tennyson_water:JPB02,Burrow:JPCA98}

A comparison o four resonance patterns with the electron transmission spectroscopy (ETS) data of Burrow and
coworkers \cite{Burrow:JPCA98,Burrow:PRL04} can be
found in Figs.
\ref{fig:comparison_uracil}-\ref{fig:comparison_guanine}, where the time-delay
data from our calculations has been rescaled by an overall constant and shifted
down in energy to
facilitate the comparison of energies, widths and spacing. All of the
resonances obtained in our calculations 
are listed in Table \ref{tab:resonances}.

\begin{table}
\begin{tabular}{|l|c|c|c|} \hline
Molecule & Energy (eV) & Width (eV) & Partial wave \\ \hline
\emph{Uracil}   &2.16  &0.2  &3 ($50\%$)  \\ 
  &5.16  &0.6  &2 ($66 \%$)  \\ 
  &7.8  &0.9  &3 ($64 \%$)  \\ \hline

\emph{Thymine}   &2.4  &0.2  &3 ($53\%$)  \\ 
  &5.5  &0.6  &2 ($62 \%$)  \\ 
  &7.9  &1.0  &3 ($61 \%$)  \\ \hline

\emph{Cytosine}   &1.7  &0.5  &3 ($51\%$)  \\ 
  &4.3  &0.7  &2 ($68 \%$)  \\ 
  &8.1  &0.8  &3 ($63 \%$)  \\ \hline

\emph{Adenine}   &2.4  &0.2  &2 ($65\%$)  \\ 
  &3.2  &0.2  &2 ($62 \%$)  \\ 
  &4.4  &0.3  &3 ($51 \%$)  \\ 
  &9.0  &0.5  &5 ($53 \%$)  \\ \hline

\emph{Guanine}   &2.4  &0.2  &2 ($46\%$)  \\ 
  &3.8  &0.25  &2 ($44 \%$)  \\ 
  &4.8  &0.35  &4 ($38 \%$)  \\ 
  &8.9  &0.6  &5 ($33 \%$)  \\ 
  &12  &1.0  &4,5 (both $23 \%$)  \\ \hline

\end {tabular}
\newpage
\caption{Energies, widths and dominant partial waves of the resonances
discussed in the text.}
\label{tab:resonances}
\end{table}

\subsubsection{Uracil}
In the cross section of uracil we find 3 resonances, at 2.16 eV (of width 0.2
eV), at 5.16 (0.6 eV wide) and a very broad resonance at 7.8 eV. 
The resonance at 2.16 eV is dominated by the $l=3$ partial wave (50\%) and has
contributions from $l=1$ (35\%) and $l=2$ (11\%). at 5.16 eV the main partial
wave is d (66\%), at 7.8 eV f-wave is the dominant contribution (64\%).

In the work of Gianturco $et$ $al.$ 
(see Ref. \onlinecite{Gianturco:uracil_JCP04})  
three $\pi^*$ resonances are found at energies of 2.2, 3.5 and 6.5 eV.
The second and
third $\pi^*$ resonances from that work fall at lower energies than 
ours, a somewhat surprising discrepancy since the theoretical models are very
similar.

The contribution of the dipole field at distances
larger than 12 Bohr is neglected in Ref. \onlinecite{Gianturco:uracil_JCP04}, but we
have noticed that this influences only the overall magnitude of the cross
sections (roughly an increase of 20\% at very low energy, that is reduced to about
5\% around 10eV), the dipole physics only weakly affects the resonance positions and widths.

Resonances are measured Ref. \onlinecite{Burrow:JPCA98} to occur at 0.3, 1.5 and
3.8 eV.
They are all assigned as $\pi$ resonances, \cite{Burrow:PRL04} so our results should be
shifted by about 2 eV down, whereas the spacing between the resonances is
larger than experiment. The relative resonance widths are similar to Ref. 
\onlinecite{Burrow:PRL04}, in that the first resonance is very narrow, the
second broader and the third very broad, a comparison is shown in Fig.
\ref{fig:comparison_uracil}, where an integration of the experimental data has
been performed to show more clearly the resonance positions and widths.


\subsubsection{Cytosine}
For cytosine we find 3 main resonances,  a very
sharp one at 1.7 eV (width 0.5 eV), then at 4.3 eV (width 0.7 eV) and a third at 8.1
eV (width 0.8 eV). The dominant angular momentum character of the resonances is
the same as for the three corresponding resonances of uracil.
Comparing the resonance positions with the data of Ref. 
\onlinecite{Burrow:JPCA98} we see the same general trend already observed with
uracil, of an overall shift higher than experiment of all resonances by about 2 eV. Interestingly, the
first two resonances are measured to occur at an energy lower than in uracil, a
trend that we  verify in our
calculations.


\subsubsection{Thymine}
The scattering cross section for thymine is  closely similar to
uracil, 
which is not surprising in view of their close structural similarities, this
applies to 
both the magnitude and the position of the resonances, which are 
slightly shifted to higher energies. Specifically, we find resonances at
2.4 eV  (width 0.2 eV) at  5.5 eV (width 0.6 eV) and at 7.9 eV (width 1 eV).

\subsubsection{Adenine}
The electron scattering spectrum for adenine presents many resonances, due to the complexity
of the target structure, as expected. Also very interesting is the
fact that the cross section drops sharply at energies below 2 eV, a behavior
opposite to that found for 
the other molecules, if we do not consider the dipole physics outside
the R-matrix box, whereas a zero-energy peak appears in the full calculation,
a possible sign of a dipole bound state right below threshold. 

The first resonance occurs at
2.4 eV (width 0.2 eV), the second at 3.2 eV (sharp, width 0.2 eV), then another
centered at 4.4 eV  (0.3 eV wide), while at
9 eV we have a  broader resonance of width 0.5 eV.
The dominant partial wave of the first two resonances is $l=2$ (65\% and 62\%
respectively).
The third resonance is $l=3$ at 51\% and $l=4$ at 33\%
The resonance at 9 eV is dominantly $l=5$ (53\%) with an $l=3$ contribution
(22\%).

Compared to experiment we have a shift of all resonances roughly 1.5 eV higher,
as in guanine, in this case the spacings are correct (about 1
eV between the first three resonances, while the fourth falls too high in
energy and it is not measured in experiment). Also the experimental widths of the
first three resonances are very similar, as in our data. A comparison with the
data of Ref. \onlinecite{Burrow:JPCA98} is shown in Fig.
\ref{fig:comparison_adenine}. 


\subsubsection{Guanine}
For guanine we find 4 resonances: at 2.4 eV (width 0.2 eV), at 3.8 eV (width 0.25eV),
a third at 4.8 eV (width 0.35 eV), then at 8.9 eV (width 0.6 eV) and a broad
resonance around 12 eV. 
Each of the first three resonances has strong
contributions from d, f, and g-waves.
At 2.4 eV the contributions are 46\% for $l=2$  and 37\% for $l=3$, for the second
resonance $l=2$ is 44\% while $l=4$ is 32\%, the third is 38\% of $l=4$
character and 35\% of $l=3$.
The resonance at 8.9 eV is 33\% h-wave, 28\% f-wave and 20\% g-wave. At 12 eV
the composition is: $l=4$ and $l=5$  equally at 23\%, while $l=6$ contributes a
further 13\%. 

Comparison to experimental data (see Fig. \ref{fig:comparison_guanine}) shows
again a shift of 1.5 eV overall, while the
resonance spacing is well reproduced, and the second and third resonances fall
at higher energies with respect to adenine, as in our calculations. Also the
widths seem close to experiment.


\subsection{Resonance molecular structures}
\label{sec:Resonances}
From the shapes and nodal structures of the resonant states it is
possible to attempt a discussion of the dissociation patterns observed
experimentally, if we consider the resonant states as being precursors for
dissociative states. Caution must be used, though, in drawing 
conclusions from this analysis, because this involves a certain degree of
speculation.  In fact, to establish once and for all the dissociation patterns
of these complicated systems, scattering calculations at many different
geometries would
have to be carried out, and the nuclear dynamics should be included, 
At present this is  computationally too expensive to contemplate.
The first two resonances observed for uracil are shown in
Figs. \ref{res:uracil1} and \ref{res:uracil2}. The quantity plotted is a projection on the molecular
plane of the eigenvector corresponding to 
the maximum eigenvalue of the time-delay matrix \cite{Greene:rev96} 
\be
Q=i  S \frac{dS \dagger}{dE},
\ee
where $S$ is the scattering matrix.
At the energy where the time delay 
of the resonance is a maximum, this eigenvector constitutes the dominant contribution to the
resonant structure, since it corresponds to the partial wave  that experiences the
maximum time delay in the scattering process. For sufficiently narrow
resonances one eigenvalue is always dominant, making the resonance analysis
much easier.
The eigenvectors of the time-delay matrix are complex, so we adopt a phase
factor such that 
the highest peak of the wavefunction is a purely real
number. We then find that  the resonance wavefunction  is real everywhere, to a good approximation
(the imaginary part is about $10^{-6}$ smaller than the value of the real part),
and we plot only the real part.
We analyze in detail only the cases of uracil and adenine, the other
pyrimidines being
very similar to the former and guanine to the latter.

The nodal patterns for uracil are very similar to the ones showed
in Ref. \onlinecite{Gianturco:uracil_JCP04}, which is not surprising since
the approximations made in that work are similar to ours, as already discussed,
therefore we show only the first two resonances. Incidentally, we
notice the close resemblance of these resonant wavefunctions to the
first virtual orbitals of uracil from a HF calculation performed with a small
basis set (6-31G*), see for example Fig. \ref{orbital:uracil2}. 
In Ref. \onlinecite{Gianturco:uracil_JCP04} also 
a very low $\sigma^*$
resonance is found at 0.012 eV. Our cross section grows
substantially at low energy. If we plot the eigenstate corresponding to the largest
eigenvalue of the time-delay matrix (as described in Sec. \ref{sec:Resonances}),
as in Fig. \ref{res:uracil1}, corresponding to this low energy range, it looks similar to Fig. 5 in Ref. \onlinecite{Gianturco:uracil_JCP04}, with the
main differences being that in our wavefunction the N$_{3}$-H bond has a node,
there is a large excess charge on N$_{3}$ and on the oxygen attached to C$_4$,
while another nodal surface cuts diagonally from C$_2$ to C$_5$. This resonance
anyway does not appear to be so relevant in the experimental
data,\cite{Burrow:PRL04} where mainly
the $\pi^*$ resonances are detected.

We can also see that there is accumulation of electronic density (the peaks of the
wavefunction) on the ring structure, and that many of the ring bonds have nodal
surfaces cutting through them, so capture in these resonant states can be
reasonably thought as leading to a fragmentation of the molecule in which the
aromatic ring is broken.
Experimental dissociation patterns are illustrated for Br-uracil in Ref.
\onlinecite{Sanche_Illenberger:JCP00}, where evidence for breaking of the ring
structure lies in the peaks at 1.6 and 3.5 eV produced by (OCN)$^-$ and other
fragments. These fragments can be generated by capture into shape resonances,
appearing in our calculations at 2.2, 5.2 and 7.8 eV respectively. In
particular there is a nodal surface in the 5.2 eV resonance that encloses the
C$_{4}$-N$_{3}$ bond, which could generate a CN$^{-}$ fragment.

Since our calculations do not take into account core-excited states or vibrations, our
results do not include
any Feshbach resonant structures. These appear to cause at
least some of the patterns observed in experiment, as in the case of
uracil,
\cite{Burrow:PRL04} and they will presumably constitute the dominant trapping pathways for
energies higher than 7 eV, where the number of electronic Feshbach resonances starts to
become very large. \cite{Tonzani:unpublished}

In the case of uracil, we looked carefully for a $\sigma^*$
resonance that might be similar to the state shown in Fig. 3 of Ref.
\onlinecite{Burrow:PRL04} around 3 eV at equilibrium geometry, a dissociative
state most likely responsible for N$_{1}$-H bond cleavage. The  Ref. \onlinecite{Burrow:PRL04}  calculation was
performed by scaling Hartree-Fock continuum orbital energies, so no information about
the width was provided.
Such a state, taking 
into account an expected shift of 2-3 eV upward in our calculations, should
have appeared at
around 6-7 eV, and it was not found. This is probably due to the fact that this
resonance is extremely broad, since it is also not seen even in experiment.
\cite{Burrow:PRL04} Moreover in calculations carried out using complex absorbing
potentials, in connection with Green's function methods,
\cite{Santra:review,Santra_Feuerbacher:JCP04} for similar systems (like
benzene \cite{Santra_Feuerbacher:priv_comm}), analogous $\sigma$ resonances were
extremely hard to detect. They became narrower (around 1eV width at equilibrium) only
when the relevant hydrogen was
substituted with a heavy atom like chlorine; this was also demonstrated experimentally in the case of
Cl-uracil in Ref. \onlinecite{Burrow:PRL04}.

For adenine there is less experimental information available to compare. In Ref.
\onlinecite{Illenberger_adenine:CPL03} it is stated that the dominant breakup
channel for
low energy (0-4eV) electron attachment leads to hydrogen atom loss, and very
prominent resonant structures are present in the range 1 to 3 eV.  If we look at
the resonance wavefunction maps in Fig. \ref{res:adenine1}-\ref{res:adenine3} we can see that there is no significant buildup of
electronic density on any of the hydrogens, consistently with the fact that the
negative charge stays on the molecular frame, and therefore there is no H$^-$
formation.

The first few unoccupied molecular orbitals that can be obtained from a 
Hartree-Fock calculation for adenine, as in the case of uracil, are
extremely similar in their nodal structures to our resonant wavefunctions,
so these shape resonances can be viewed quite reasonably as the trapping of the
scattered electron in a virtual orbital.

Most of the C-C and C-N bonds have nodal surfaces passing through them,
This might suggest that other channels that
involve the breakup of C-C and C-N bonds could also be available at these
energies, although
probably they are less important than the hydrogen loss products.

\section{Conclusions}
\label{sec:Conclusions}

We have presented results for electron scattering from DNA and RNA bases.  In
showing some of their resonant wavefunctions, we have attempted to link 
the  
molecular breakup patterns and products to the structure of the nodal surfaces of these
wavefunctions.
The results for cross sections and resonances show an overall shift of about
1.5-2
eV higher for all the resonances in our calculations, compared to experiment,
we believe that this shift
is due to the approximate nature of our model. Apart from this, though,
we seem to reproduce trends observed in experiment, with respect to the resonance
spacing, their widths and also in relative positions for different molecules,
which gives us guarded confidence in our results. We have presented the first  
calculations of cross sections for all the main DNA
bases, and discussed the relationship of our results to experimental
data. Ample room exists for improvement in our model, and it will  be
desirable to eventually calculate resonant surfaces, and not just equilibrium
values, in order to characterize the dissociation processes. 
Work is presently underway to meet some of these challenges.

\section*{Acknowledgments}
This work was supported in part by the Department of Energy, Office of Science, by
an allocation of NERSC supercomputing resources, and by the Keck Foundation
through computational resources.  We would like to thank P. Burrow, R.
Santra and S. Feuerbacher for useful discussions and for sharing their unpublished data.

\bibliography{paper_DNA_bases}


\begin{figure}
\newpage
\newpage
\begin{spacing}{2}
\caption{Ground state equilibrium structures of the molecules considered in this paper. The black atoms
are oxygens, the dark gray circles represent carbons, the light gray atoms are
nitrogens while the small circles
are hydrogens.}\label{molec:structure}
\end{spacing}
\end{figure}

\begin{figure}
\newpage
\centerline{\includegraphics[width=18.0cm]{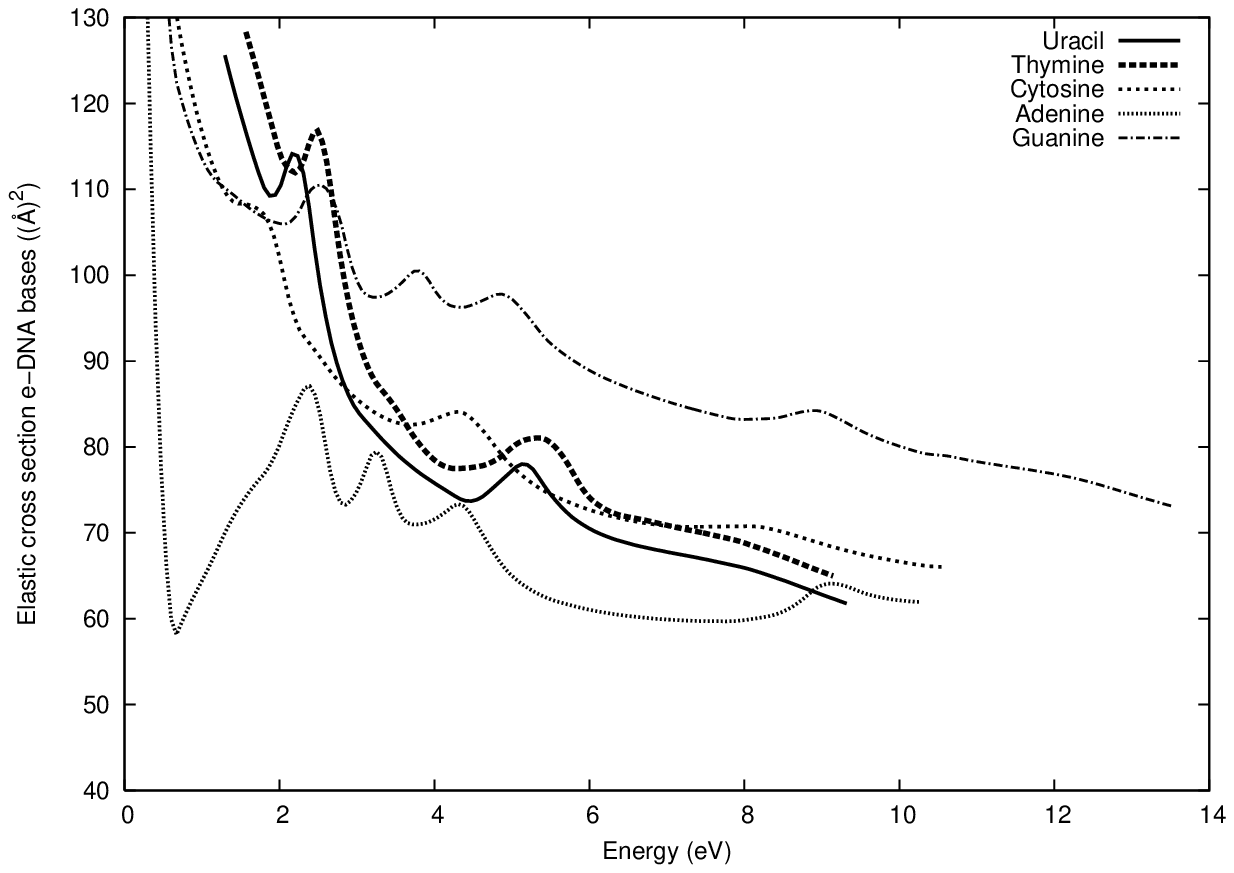}}

\begin{spacing}{2}
\newpage
\caption{Partial elastic cross section for the 5 DNA and RNA bases described in
the text. Calculations involve partial waves up to $l=10$ and the dipole physics
outside the R-matrix box is included exactly.}\label{fig:total_cross_section}
\end{spacing}
\end{figure}

\begin{figure}
\newpage
\centerline{\includegraphics[width=18.5cm]{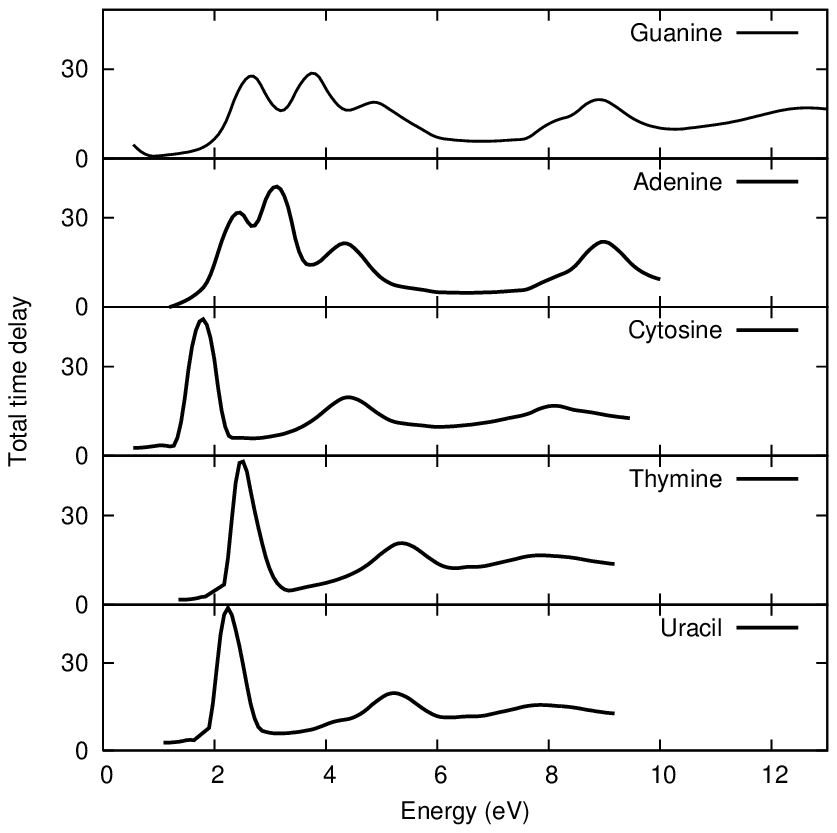}}
\newpage
\begin{spacing}{2}
\caption{ Total time-delay for the molecules described in the text.
}\label{fig:deriv_eigenphase_sum}
\end{spacing}
\end{figure}

%





\begin{figure}
\centering
\newpage
\includegraphics[width=17.0cm]{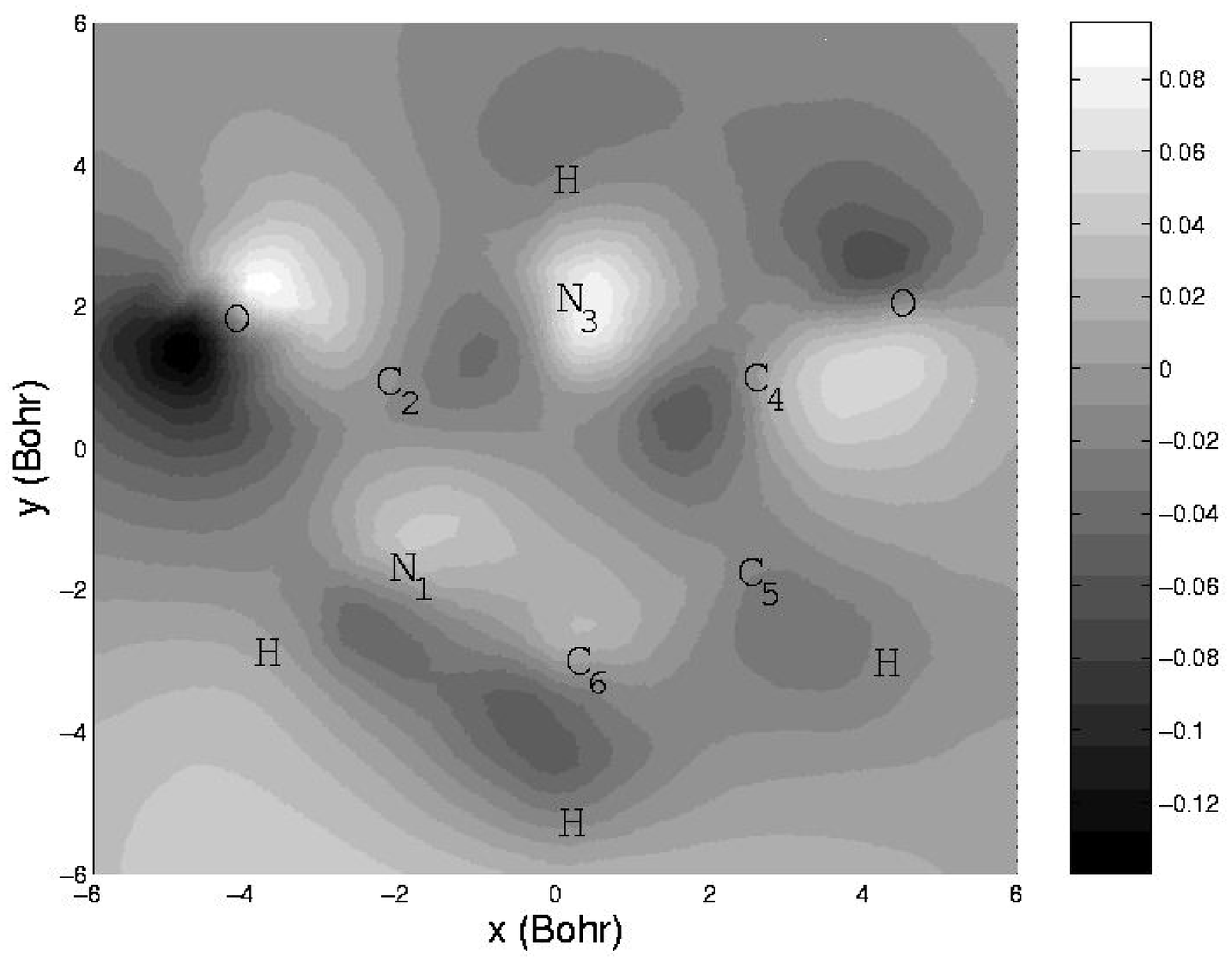}
\newpage
\caption{Uracil: 1st  resonance wavefunction. Eigenvector associated to the
dominant eigenvalue of the time-delay matrix, as described in the text, for 
scattering at 0.2 eV.}
\label{res:uracil1}


\newpage
\includegraphics[width=17.0cm]{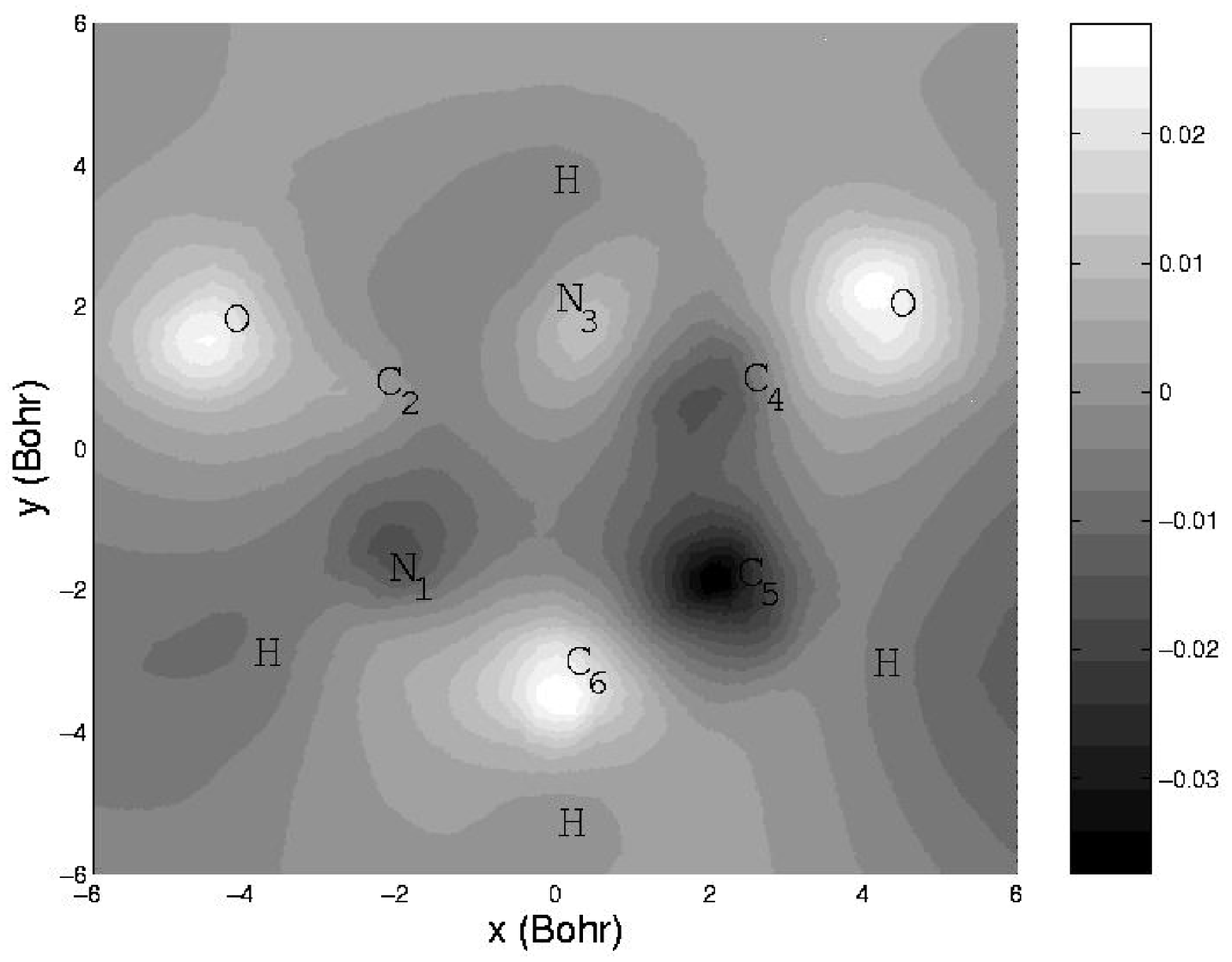}
\newpage
\caption{Uracil: 2nd resonance wavefunction. Eigenvector associated to the
dominant eigenvalue of the time-delay matrix, as described in the text, for the
resonance at 2.2 eV.}
\label{res:uracil2}


\end{figure}

\begin{figure}
\newpage
\includegraphics[width=15.0cm]{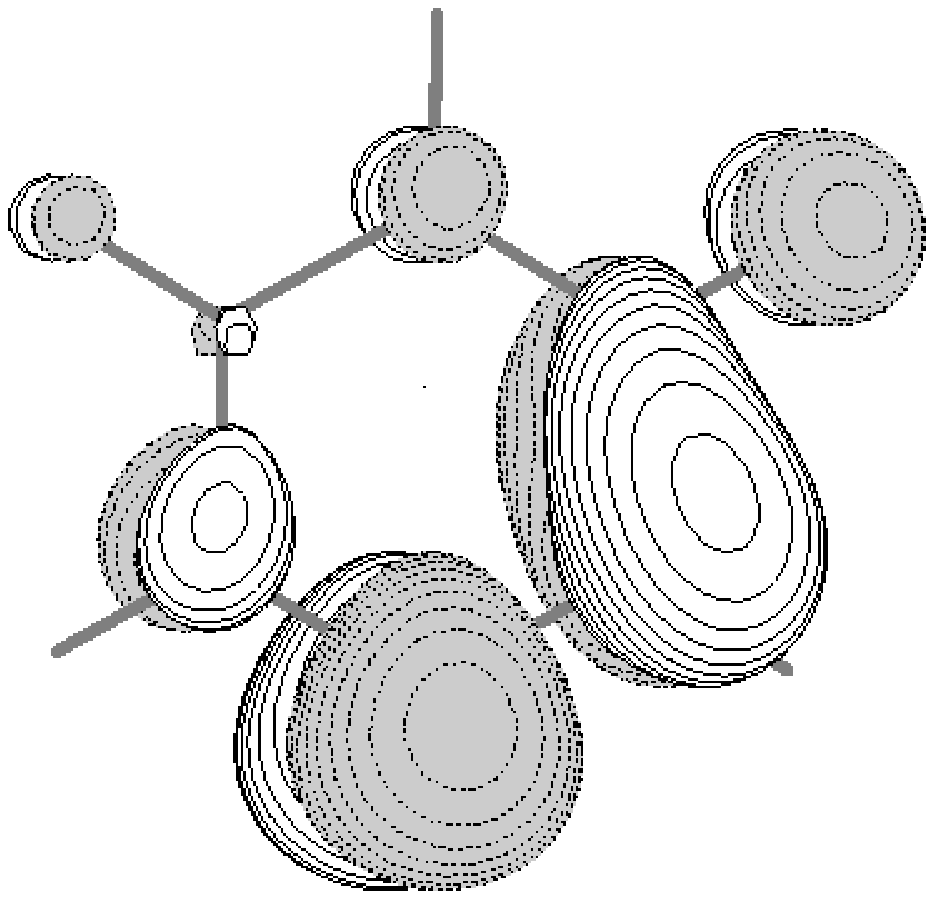}
\newpage
\caption{Uracil: structure of the virtual orbital associated to the resonance
at 2.2 eV. The energy of this virtual orbital is 3.42 eV when using a 6-31G**
basis set. The orientation of the molecule  is the same as in the previous
plots.
The black and white lobes
correspond to opposite signs.  It is possible to notice the node in the
molecular plane that makes this a $\pi^*$ orbital.}
\label{orbital:uracil2}


\end{figure}

\begin{figure}
\centering
\newpage
\includegraphics[width=17.0cm]{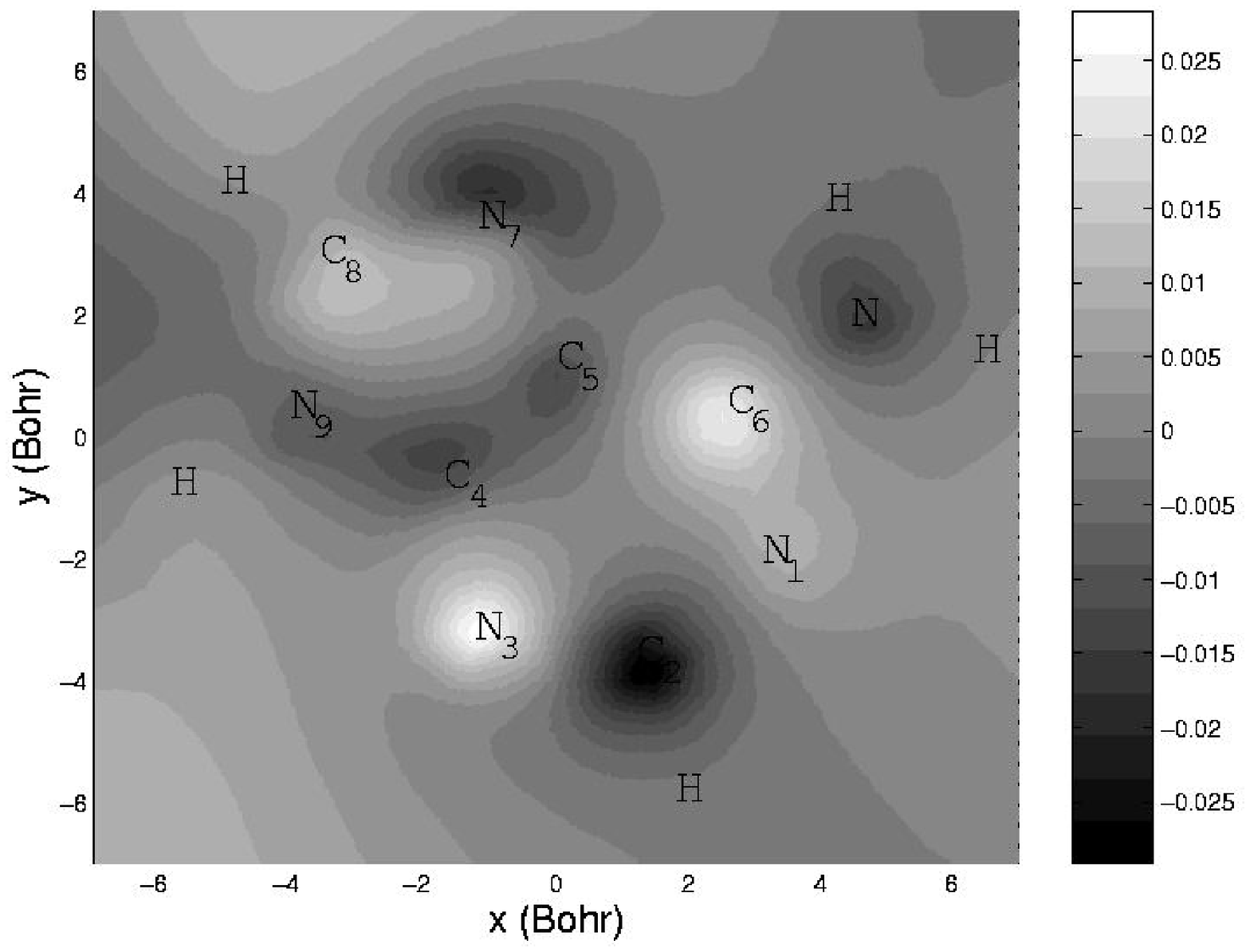}
\newpage
\caption{Adenine: 1st  resonance wavefunction. Eigenvector associated to the
dominant eigenvalue of the time-delay matrix, as described in the text,
at 2.4 eV.}
\label{res:adenine1}


\newpage
\includegraphics[width=17.0cm]{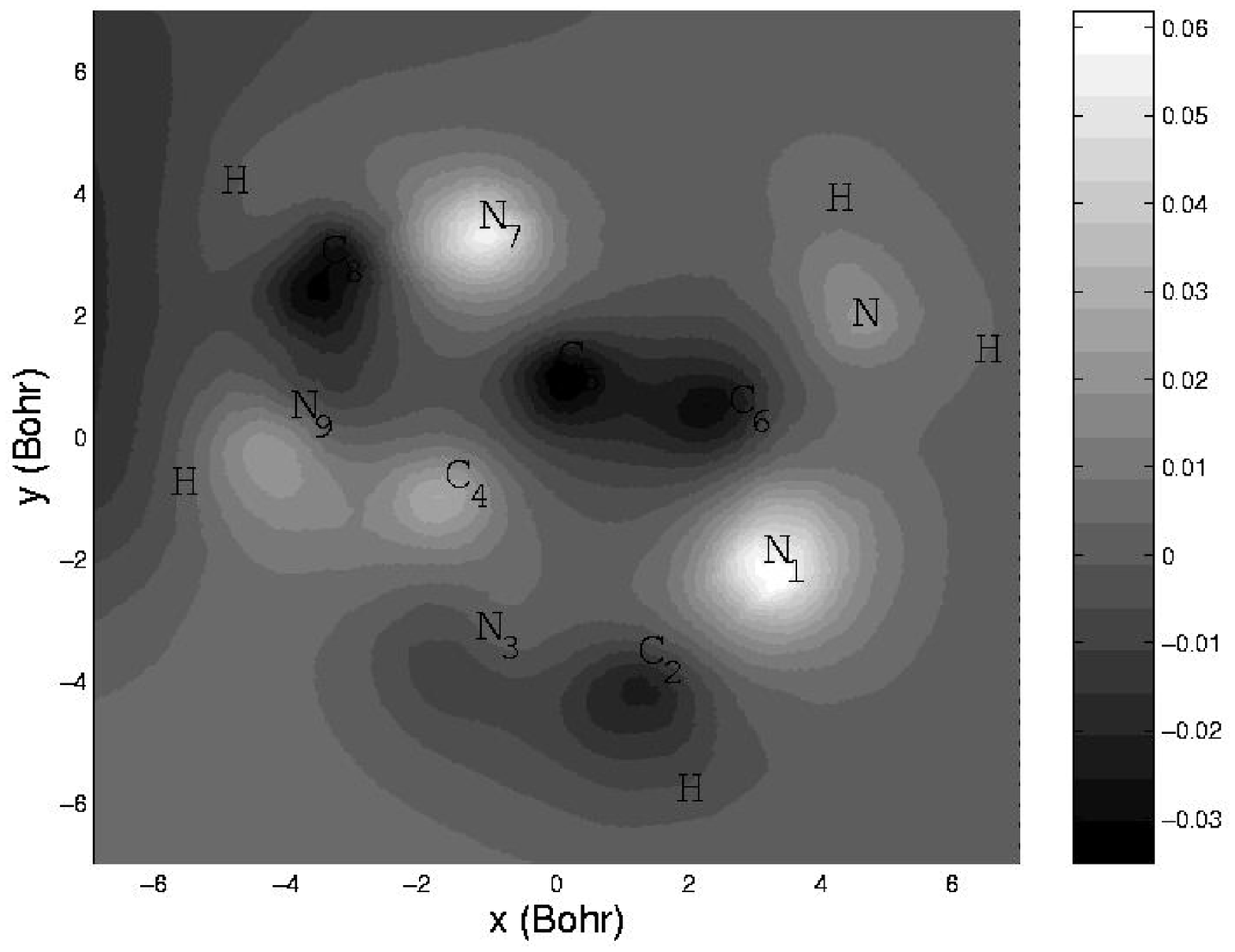}
\newpage
\caption{Adenine: 2nd resonance wavefunction. Eigenvector associated to the
dominant eigenvalue of the time-delay matrix, as described in the text, for the
resonance at 3.2 eV.}
\label{res:adenine2}


\end{figure}

\begin{figure}
\centering

\newpage
\newpage
\caption{Adenine: 3nd resonance wavefunction. Eigenvector associated to the
dominant eigenvalue of the time-delay matrix, as described in the text, for the
resonance at 4.4 eV.}
\label{res:adenine3}


\end{figure}

\begin{figure}
\newpage
\centerline{\includegraphics[width=18.5cm]{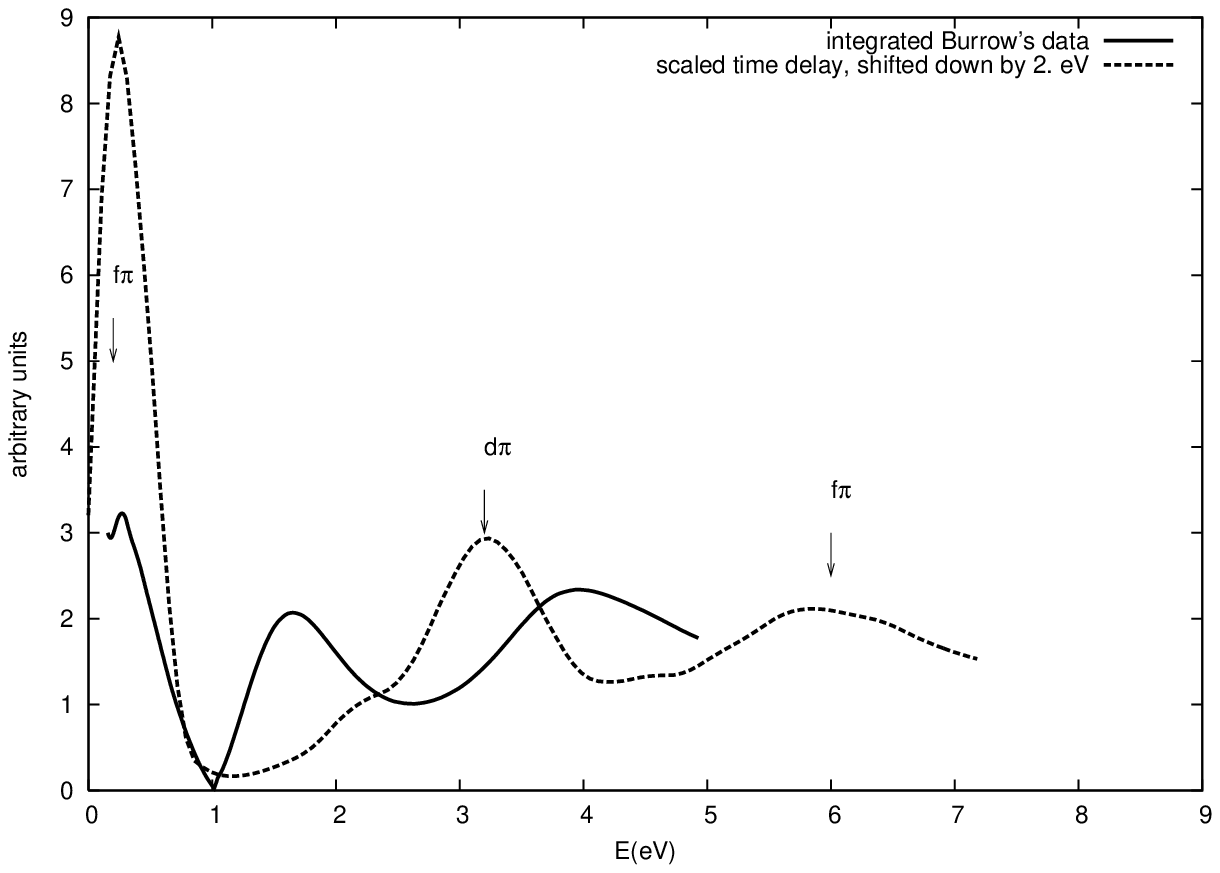}}
\newpage
\begin{spacing}{2}
\caption{Comparison with experimental data of Burrow $et$ $al.$
\cite{Burrow:PRL04} for uracil. The arrows indicate resonance positions from
the present work, while labels show the dominant partial wave of resonance. The
time-delay curve is shifted downward by 2.0 eV to have the position of the first resonance
coincide with experimental data.
}\label{fig:comparison_uracil}
\end{spacing}

\end{figure}
\begin{figure}
\newpage
\centerline{\includegraphics[width=18.5cm]{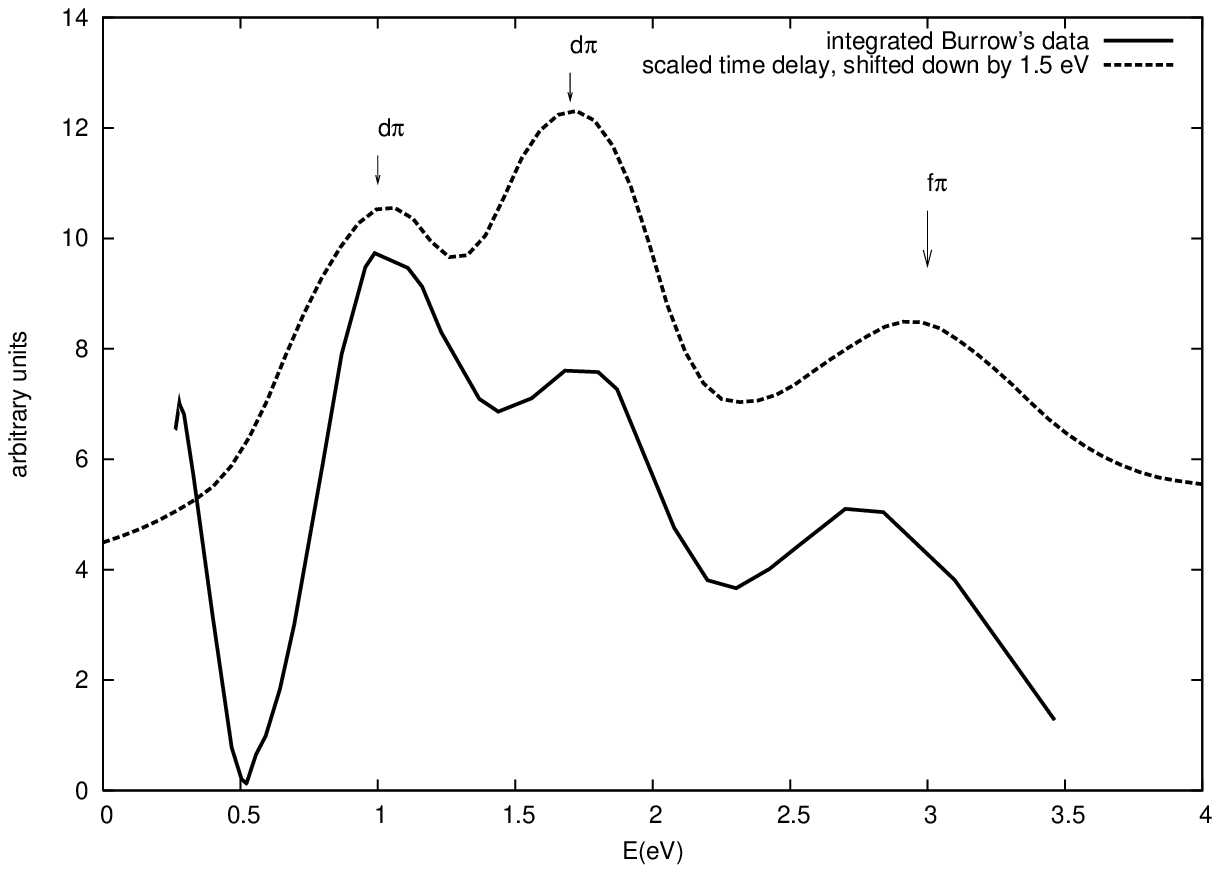}}
\newpage
\begin{spacing}{2}
\caption{Comparison with experimental data of Burrow $et$ $al.$
\cite{Burrow:JPCA98} for adenine. The arrows indicate resonance positions from
the present work, while labels show the dominant partial wave of resonance.  The
time-delay curve is shifted downward by 1.5 eV to have the position of the first resonance
coincide with experimental data.
 }\label{fig:comparison_adenine}
\end{spacing}

\end{figure}
\begin{figure}
\newpage
\centerline{\includegraphics[width=18.5cm]{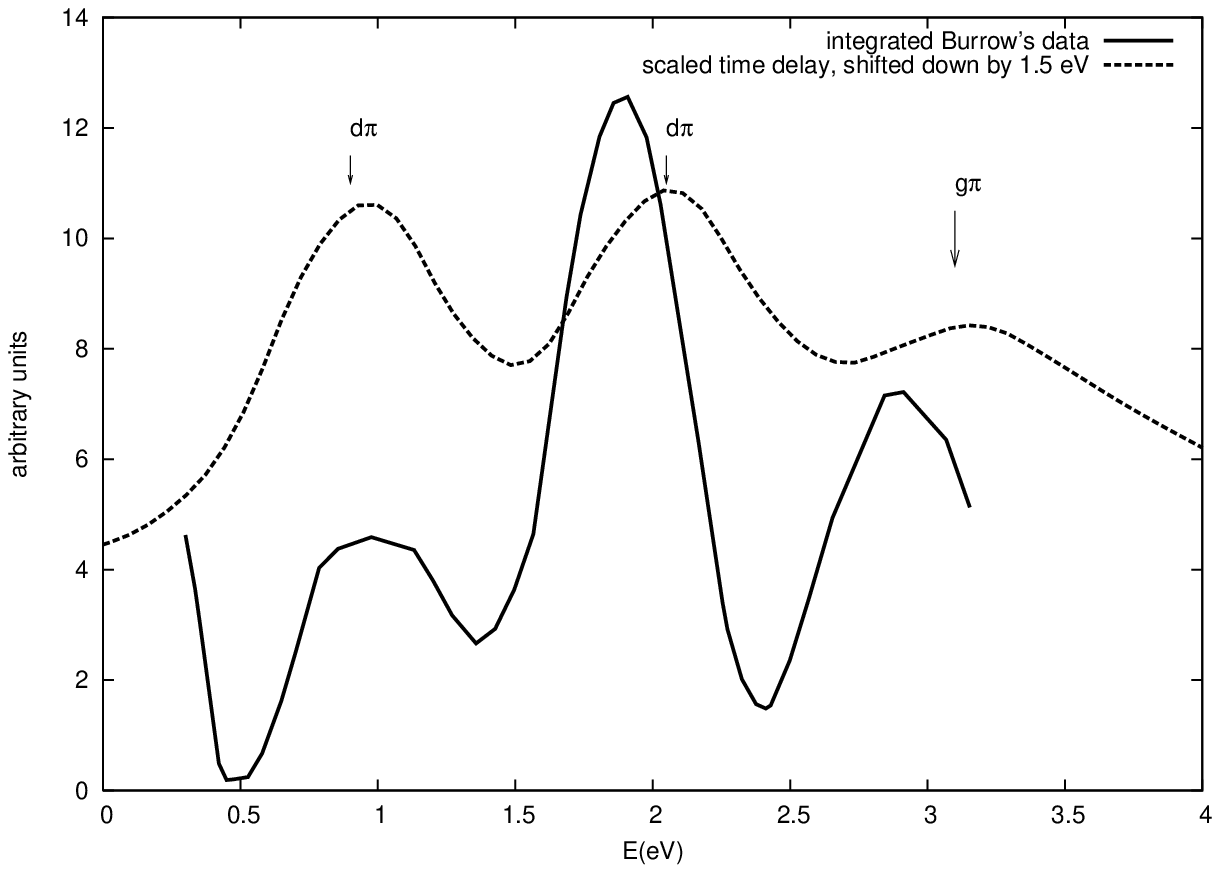}}
\newpage
\begin{spacing}{2}
\caption{Comparison with experimental data of Burrow $et$ $al.$
\cite{Burrow:JPCA98} for guanine. The arrows indicate resonance positions from
the present work, while labels show the dominant partial wave of resonance. The
time-delay curve is shifted downward by 1.5 eV to have the position of the first resonance
coincide with experimental data.
 }\label{fig:comparison_guanine}
\end{spacing}

\end{figure}

\end{spacing}
\end{document}